\pdfoutput=1

\documentclass[
    ,final           
  ]
  {aipproc}

\layoutstyle{6x9}

\begin{document}

\title{Forward Physics at CMS}

\classification{12.38.Aw - 13.85.Hd - 13.87.Ce}
\keywords      {Forward Energy Flow - Forward Jets}

\author{Beno\^it Roland, on behalf of the CMS Collaboration}{
  address={University of Antwerp, Campus Groenenborger, 
2020 Antwerp, Belgium}
}

\begin{abstract}
 
Studies of forward processes are important tests of the standard model and inputs for Monte Carlo tuning. 
A measurement of the energy flow in the forward pseudorapidity region of CMS, $3.15 < \vert \eta \vert < 4.9$, 
is presented for 3 values of the centre-of-mass energy $\sqrt{s}$ = 0.9 TeV, 2.36 TeV and 7 TeV. The forward energy flow 
is measured for minimum bias events and for events with a central dijet system the transverse energy of which provides 
a hard scale. The energy flow is compared to various Monte Carlo models with different multiparton interaction schemes. 
A study of forward jets in the pseudorapidity range $3.2 < \vert \eta \vert < 4.7$ is presented for $\sqrt{s}$ = 7 TeV. 
\end{abstract}

\maketitle

\section{Measurement of the forward energy flow}
Measurements in the forward region probe the parton content of the proton at small values of the
proton momentum fraction, in a region where the parton densities might become very large and where the probability 
for more than one partonic interaction per event should increase. 
The measurement of the energy flow in the CMS forward region \cite{FwdEFlow:2010}, in the pseudorapidity 
range $3.15 < |\eta| < 4.9$, should therefore be sensitive to the modelling of parton radiation at large $\eta$ 
and the description of multiparton interactions (MPI) \cite{Sjostrand:1987su}.
The measurement is performed for two event classes: a minimum bias sample and a sample of events 
with a central dijet system. The first class is characterized by
zero or few partonic interactions, while the second one has at least one hard scattering at the parton level. 
The amount of parton radiation in the forward region is therefore expected to be larger for the dijet sample, 
which should be directly reflected in the forward energy flow. 
The forward measurement can thus provide a complementary and independent constraint on the MPI modelling.
\subsection{HF calorimeter and Trigger subsystem}
A detailed description of the CMS experiment can be found elsewhere~\cite{:2008zzk} and we only describe here
the subsystems used to obtain the presented results. The two Hadronic Forward calorimeters HF+ and HF-,
located at $\pm$11.2 m from the nominal interaction point (IP), cover the pseudorapidity region 
$2.9 < |\eta| < 5.2$. These are Cerenkov calorimeters made of radiation hard quartz fibers embedded into 
steel absorbers.
Two subsystems, the Beam Scintillator Counters (BSC) 
and the Beam Pick-up Timing for the eXperiments (BPTX) are used to trigger the detector readout \cite{BSCBPTX}. 
The two BSCs are located at $\pm$10.86~m from the IP and cover 
the pseudorapidity region $3.23 < |\eta| < 4.65$. Each is a set of 16 scintillator tiles. The BSC elements 
have a time resolution of 3 ns and are designed to provide hit and coincidence rates. The two BPTXs, located around 
the beam pipe at $\pm$175 m from the IP, are designed to provide precise information on the bunch structure and timing
of the incoming beam, with better than 0.2 ns time resolution.  
\subsection{Analysis strategy and event selection}
The forward energy flow is measured at 3 values of $\sqrt{s}$ = 0.9 TeV, 2.36 TeV and 7~TeV, 
for the 2 event classes described above. The following conditions are imposed to select 
the minimum bias sample. A signal is required in each of the BSCs in conjunction with BPTX signals 
from both beams passing the IP. This condition rejects a large fraction of diffractive events. 
A primary vertex is required with $|z| < $ 15 cm and a transverse distance from the $z$ axis smaller than 2 cm. 
Further cuts are applied to reject beam-halo event candidates, beam-scraping events and events 
with large signals consistent with noise in HF. 
The energy flow is then measured in 5 different $\eta$ bins by summing up  
all the energy deposits in the HF towers above a noise threshold of 4~GeV. 
The following conditions are imposed to select the dijet sample. Jets are reconstructed by means of the anti-$k_T$ 
jet algorithm \cite{Cacciari:2008gp} with R = 0.5. The dijet sample consists of events with at least two leading
jets with $|\eta| < 2.5$ and $|\Delta \phi(j_1,j_2) - \pi| < 1$. Jets are required to have $p_T > 8$~GeV
at $\sqrt{s}$ = 0.9 TeV and 2.36 TeV, and $p_T > 20$ GeV at $\sqrt{s}$ =  7~TeV.
\subsection{Results and Monte Carlo Comparison}
\begin{figure}[htb] 
\begin{minipage}[t]{.32\textwidth}
\centerline{\includegraphics[angle=90,width=1.\textwidth]{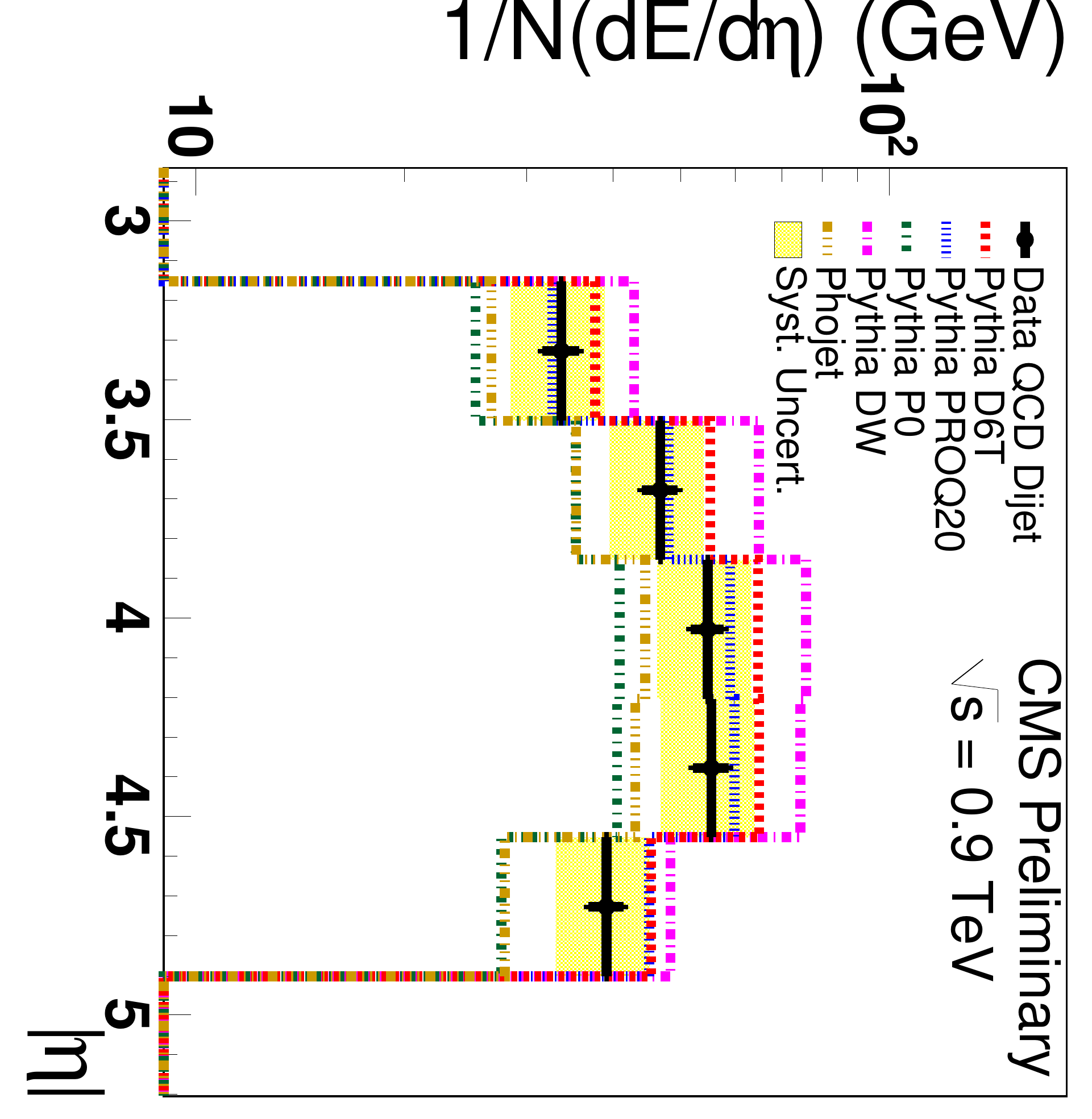}}
\end{minipage}
\begin{minipage}[t]{.32\textwidth}
\centerline{\includegraphics[angle=90,width=1.\textwidth]{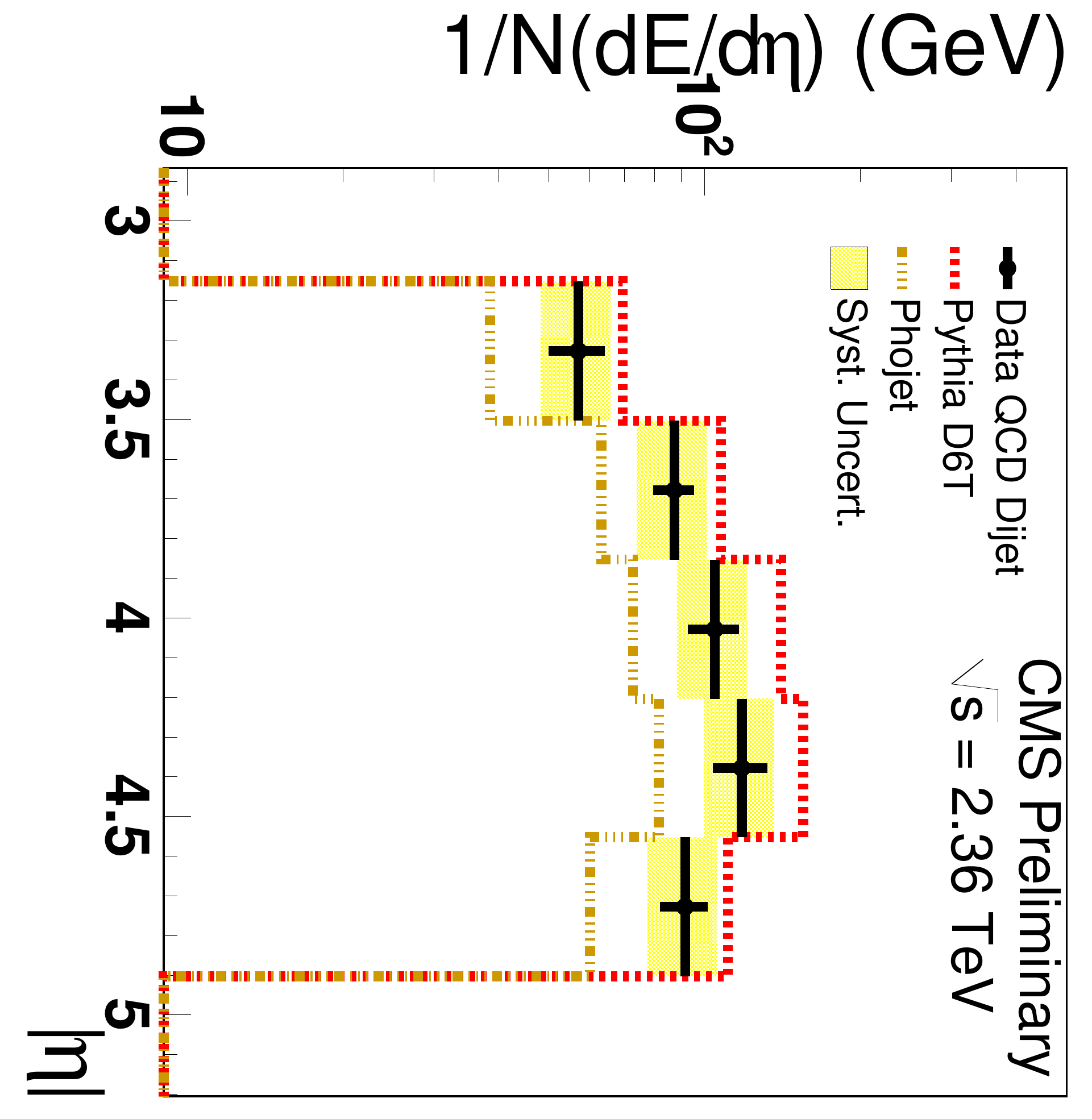}}
\end{minipage}
\begin{minipage}[t]{.32\textwidth}
\centerline{\includegraphics[angle=90,width=1.\textwidth]{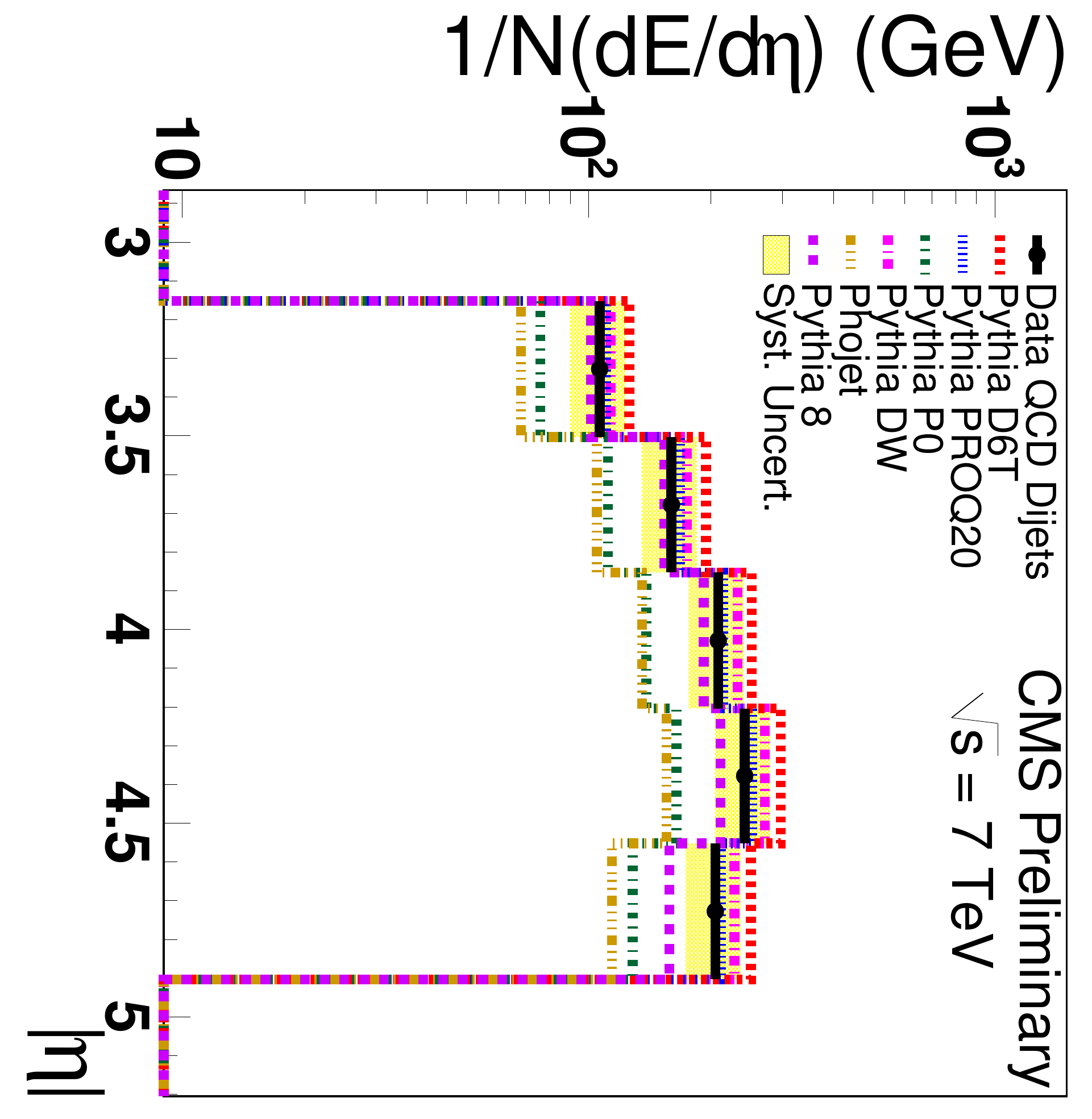}}
\end{minipage}
\caption{Energy flow in the minimum bias sample (top) and dijet sample (bottom) as a function of $\eta$
at $\sqrt{s}$ = 0.9 TeV (left) , 2.36 TeV (center) and 7~TeV (right). Detector level distributions are compared to 
predictions from the Monte Carlo event generators PYTHIA and PHOJET. Error bars correspond to 
statistical errors. The dashed bands represent the total systematic uncertainty of the measurement.}
\label{fig:EflowMeasurement}
\end{figure}

Figure \ref{fig:EflowMeasurement} shows the energy flow in HF at detector level, in the region $3.15 < |\eta| < 4.9$,
for the minimum bias and dijet samples, at $\sqrt{s}$ = 0.9 TeV, 2.36 TeV and 7~TeV. 
Error bars indicate statistical uncertainties, while the dashed bands represent the total systematic uncertainty.
The dominant contribution of 15 $\%$ corresponds to the global HF energy scale uncertainty.  
Detector level distributions are compared to predictions from Monte Carlo event generators 
passed through the full CMS detector simulation based on Geant4 \cite{Agostinelli:2002hh}.
The predictions from PYTHIA6 \cite{Sjostrand:2006za} use different sets of parameters for the MPI:
the tunes D6T \cite{Field:2009zz}, DW \cite{Field:2009zz}, PROQ20 \cite{Buckley:2009bj} 
and Perugia P0 \cite{Skands:2009zm}. In the latter case a new MPI model is used \cite{Wicke:2008iz}.
The predictions from the Monte Carlo event generators PYTHIA8 \cite{Sjostrand:2007gs} 
and PHOJET \cite{Bopp:1998rc} are also shown. The forward energy flow in the miminum bias sample shows 
a stronger energy dependence in data than in Monte Carlo. This behaviour is not observed in the dijet event sample. 
The energy flow in the 900 GeV minimum bias sample is best
described by the D6T tune, whereas the PROQ20 and P0 tunes and PHOJET underestimate the data. At $\sqrt{s}$ = 7~TeV,
all the Monte Carlo predictions underestimate the data, with PYTHIA8 being close to the PROQ20 tune.
In the dijet sample, the D6T tune is too high compared to the data, the P0 tune and PHOJET are too low, 
and the best description is given by the PROQ20 tune and PYTHIA8.
\begin{figure}[htb] 
\begin{minipage}[t]{.32\textwidth}
\centerline{\includegraphics[angle=90,width=1.\textwidth]{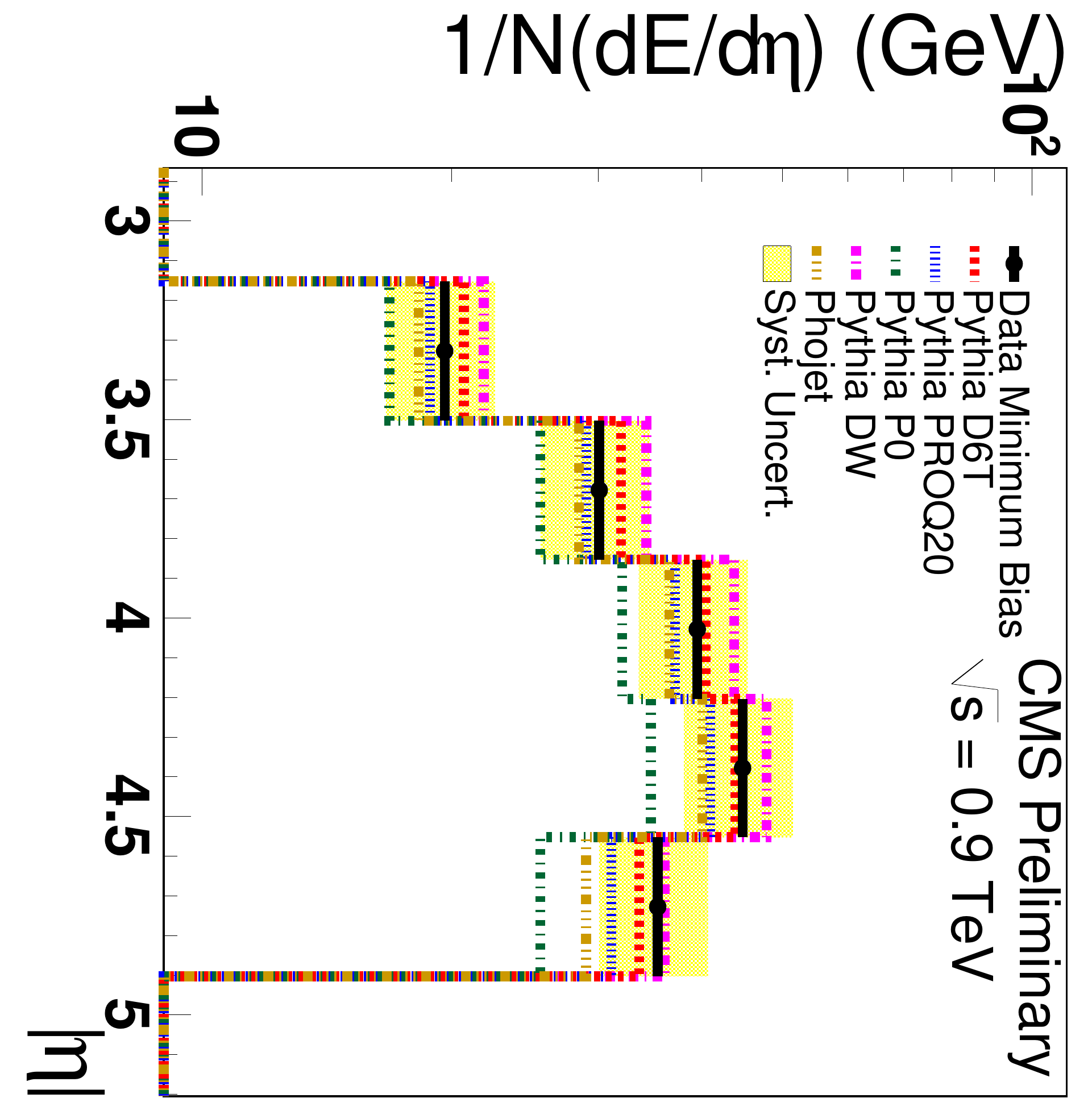}}
\end{minipage}
\begin{minipage}[t]{.32\textwidth}
\centerline{\includegraphics[angle=90,width=1.\textwidth]{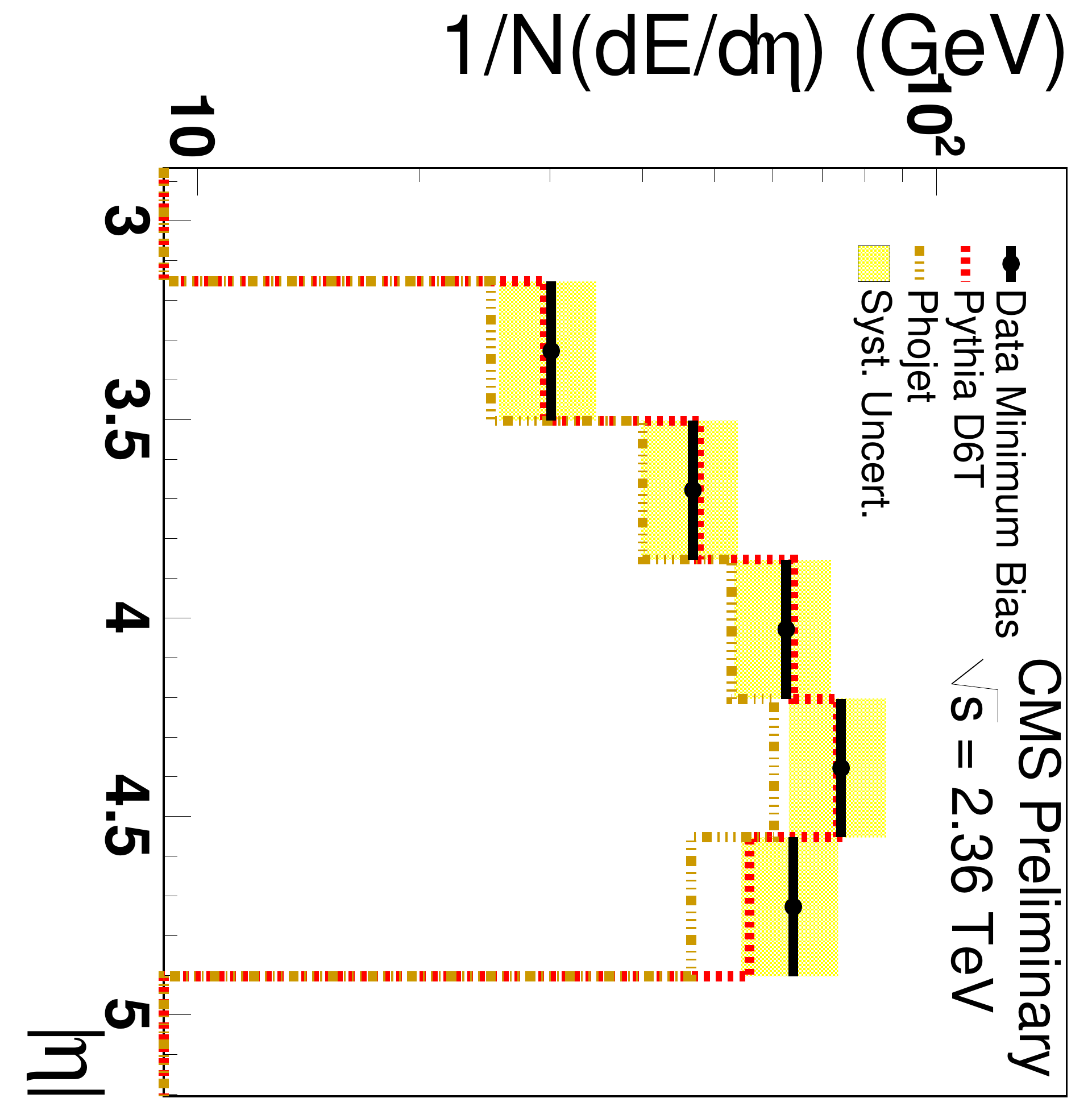}}
\end{minipage}
\begin{minipage}[t]{.32\textwidth}
\centerline{\includegraphics[angle=90,width=1.\textwidth,height=4.6cm]{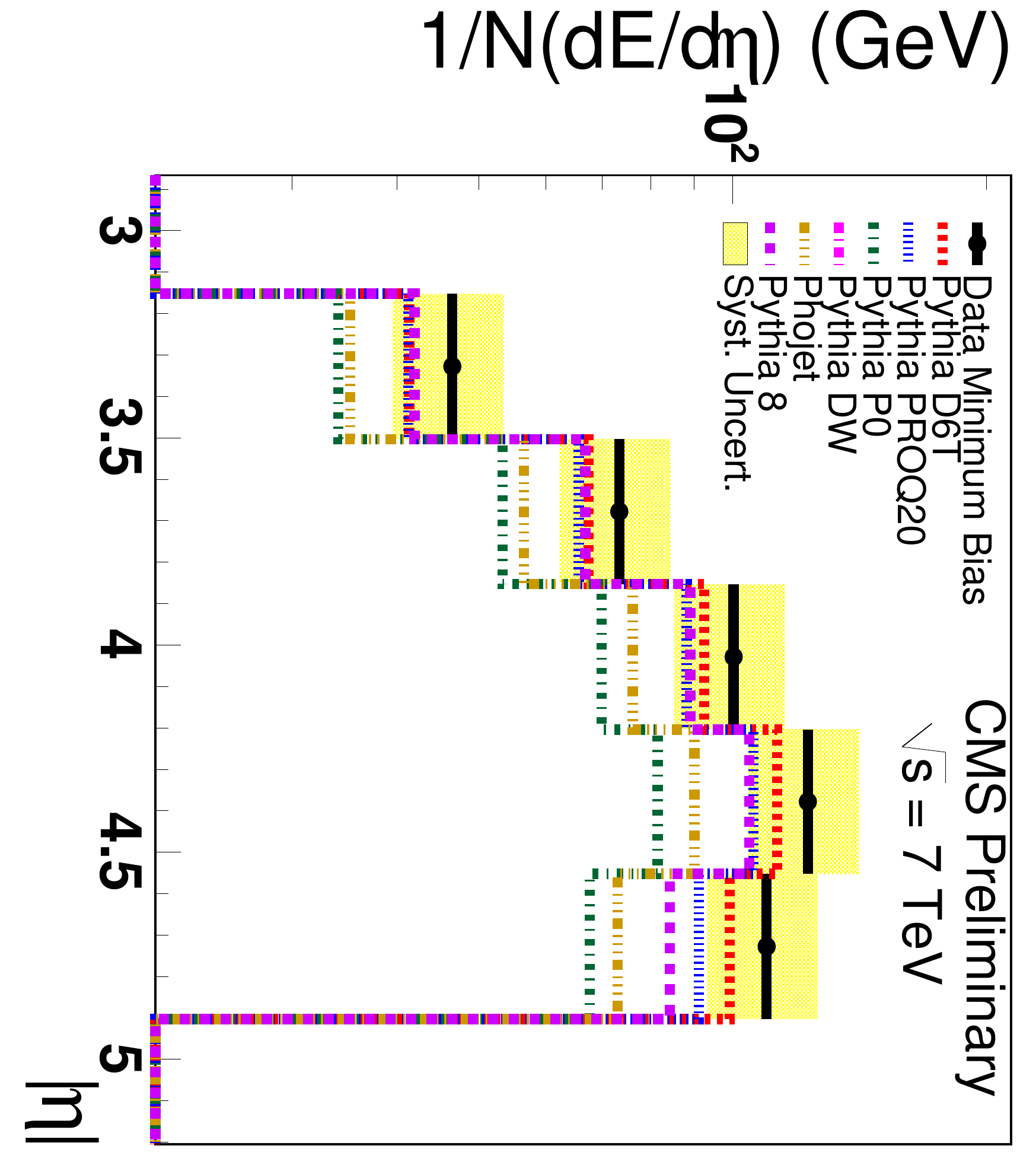}}
\end{minipage}
\end{figure} 
\vspace{-0.4cm}
\vspace{-0.4cm}
\section{Reconstruction of forward jets}
\begin{figure}[htb] 
\begin{minipage}[t]{0.33\textwidth}
\centerline{\includegraphics[angle=90,width=0.9\textwidth]{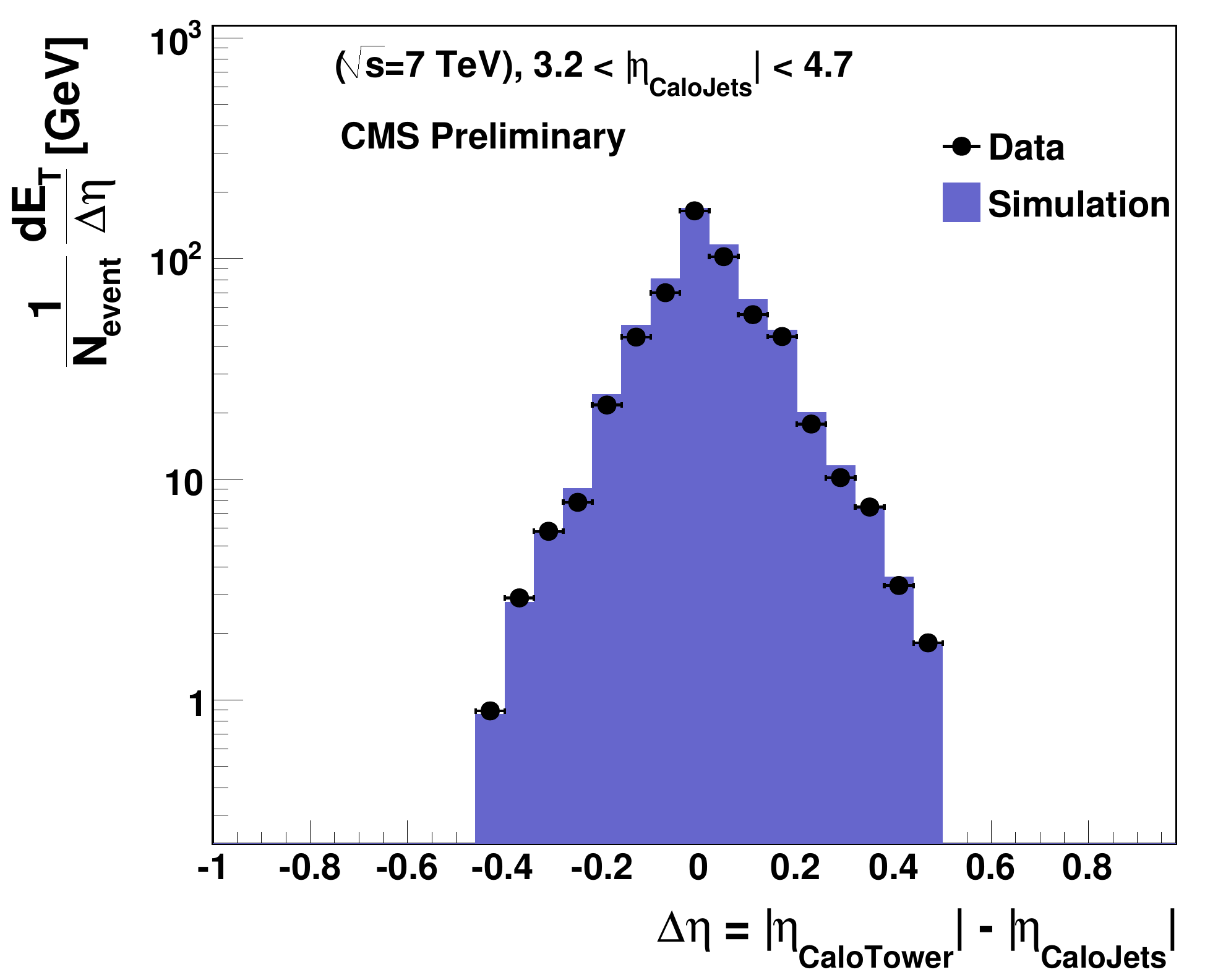}}
\end{minipage}
\begin{minipage}[t]{0.33\textwidth}
\centerline{\includegraphics[angle=90,width=1.0\textwidth]{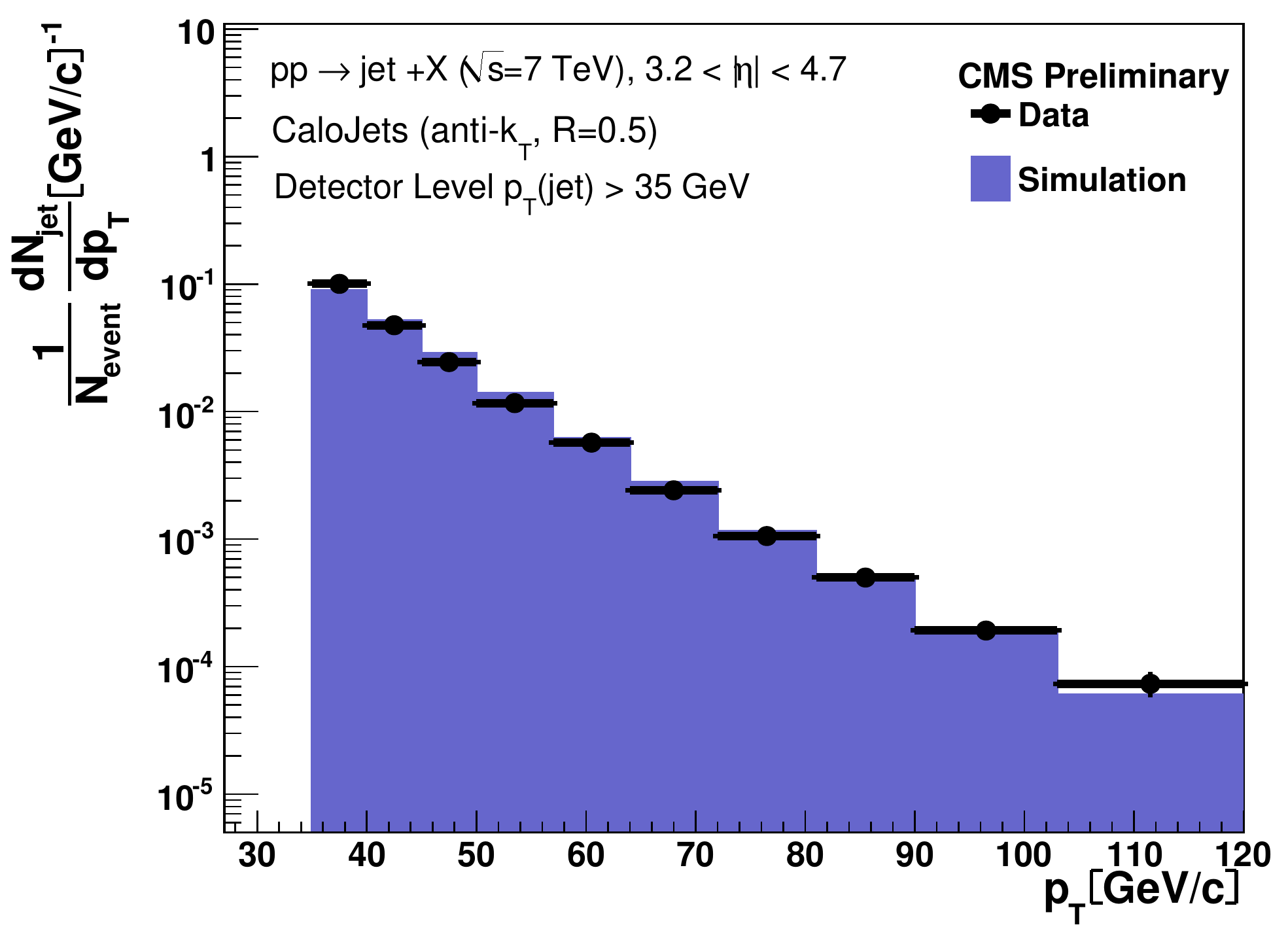}}
\end{minipage}
\begin{minipage}[t]{0.33\textwidth}
\centerline{\includegraphics[angle=90,width=1.1\textwidth]{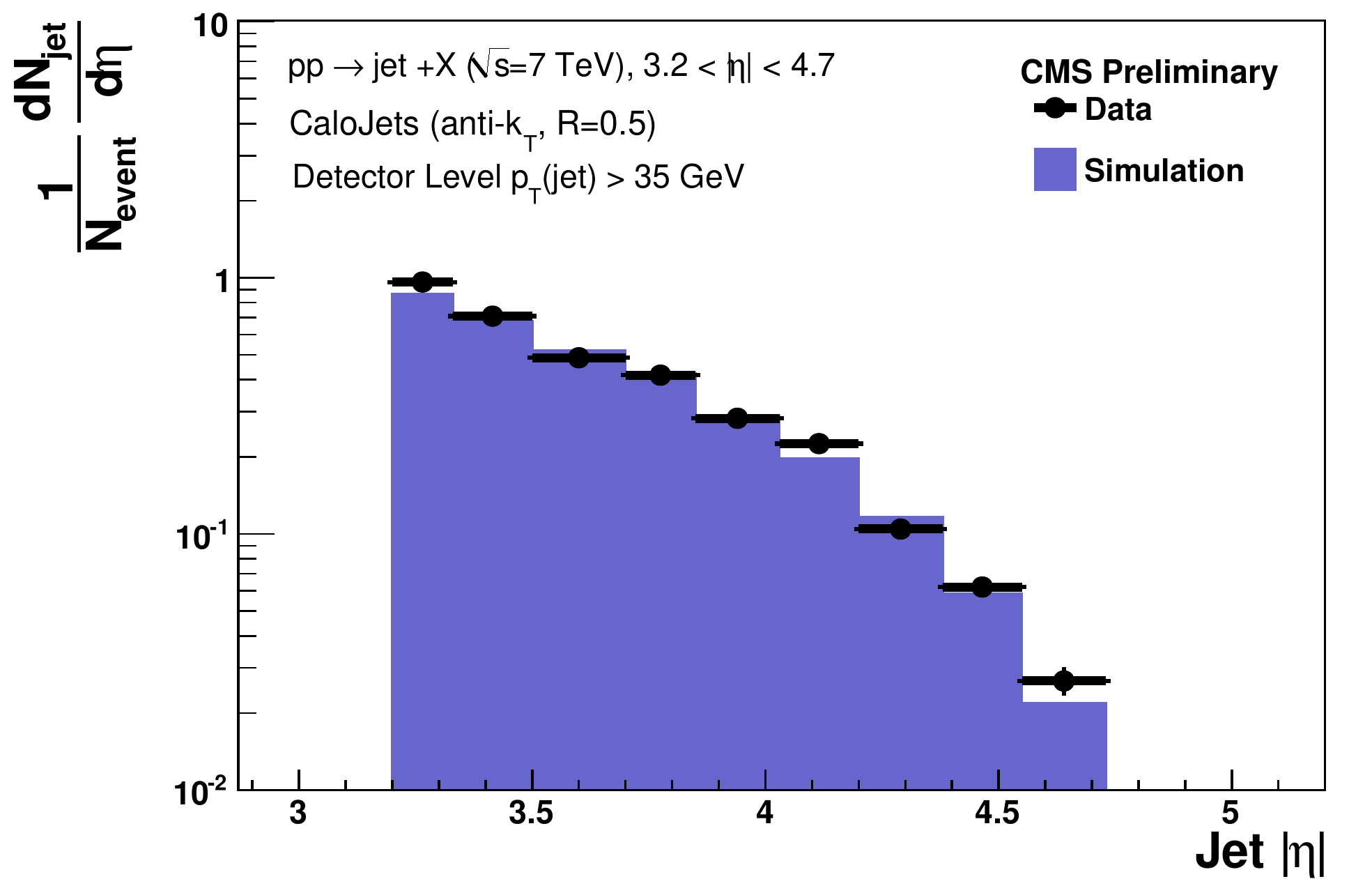}}
\end{minipage}
\caption{The transverse energy flow $E_T$ inside of the jets reconstructed in HF as a function of the distance
$\Delta \eta$ with respect to the jet axis (left) . The forward jet $p_T$ spectrum 
(center) and $\eta$ spectrum (right). Jets are required to have $35 < E_T < 120$ GeV 
and $3.2 < |\eta| < 4.7$. Detector level distributions are compared to predictions from the Monte Carlo event generator
PYTHIA6 using the D6T tune.}
\label{fig:FwdJets}
\end{figure}

Jet production has never been investigated at hadron colliders in such a forward region as the one covered
by the HF calorimeter. The first step before measuring forward jets in HF is to validate the jet reconstruction
in that region \cite{FwdJets:2010}. The event selection is similar to that used to select the miminum
bias sample. Jets are reconstructed by means of the anti-$k_T$ jet algorithm \cite{Cacciari:2008gp} with R = 0.5,
are required to have $35 < E_T < 120$ GeV and $3.2 < |\eta| < 4.7$. The jet energy is corrected for energy loss
and non-linear response of the calorimeter. The sample used for the study has been recorded at $\sqrt{s}$ = 7 TeV
and corresponds to an integrated luminosity of 10 nb$^{-1}$. 
Figure \ref{fig:FwdJets} shows the distribution of the transverse energy flow $E_T$ inside of the jets 
as a function of the distance $\Delta \eta$ with respect to the jet axis (left), the forward jet $p_T$ spectrum 
(center) and $\eta$ spectrum (right). Detector level distributions with statistical errors only are compared 
to the predictions of the Monte Carlo event generator PYTHIA6 using the D6T tune. 
A reasonable agreement is found for the different distributions.
\vspace{-0.3cm}
 \section{Conclusion}
The first measurement at hadron colliders of the forward energy flow in the region $3.15 < |\eta| < 4.9$ has been
presented, for a minimum bias sample and a sample of events with a hard scale defined by a central dijet system.
The forward energy flow in the miminum bias sample shows a stronger energy dependence in data than in Monte Carlo. 
At $\sqrt{s}$ = 7~TeV, all the Monte Carlo predictions underestimate the measured minimum bias energy flow. 
Such a behaviour is not observed in the dijet event sample. The Monte Carlo tunes giving the best description in the
forward region differ from those giving the best description of the charged particle spectra 
in the central region \cite{Krajczar:2009zz}. The validation of the forward jets reconstruction 
in the HF acceptance has been presented at $\sqrt{s}$ = 7~TeV.

\vspace{-0.3cm}
\begin{theacknowledgments}
I would like to thank my colleagues of the CMS Forward group for the
discussions and for providing the results presented in these proceedings.

\end{theacknowledgments}

\bibliographystyle{aipproc}

\end{document}